\documentclass[preprint2]{aastex}

\newcommand{\kms}{km s$^{-1}$}
\newcommand{\msun}{$M_{\odot}$}
\newcommand{\water}{H$_{2}$O}
\newcommand{\methanol}{CH$_{3}$OH}
\newcommand{\mfour}{CH$_{3}$OH $7_{0}\ \rightarrow\ 6_{1}\ A^{+}$}
\newcommand{\mnine}{CH$_{3}$OH $8_{0}\ \rightarrow\ 7_{1}\ A^{+}$}
\newcommand{\mone}{CH$_{3}$OH $6_{-1}\ \rightarrow\ 5_{0}\ E$}

\shorttitle{Maser Emission around Low-Mass Protostars}
\shortauthors{Kang et al.}
\slugcomment{To appear in ApJS}

\begin{document}
\fontsize{10}{10.6}\selectfont

\title{Water and Methanol Maser Survey of Protostars
       in the Orion Molecular Cloud Complex}

\author{\sc Miju Kang$^1$, Jeong-Eun Lee$^2$, Minho Choi$^1$,
            Yunhee Choi$^{3,4}$, Kee-Tae Kim$^1$,
            James Di Francesco$^{5,6}$, and Yong-Sun Park$^7$}
\affil{$^1$ Korea Astronomy and Space Science Institute,
            776 Daedeokdaero, Yuseong, Daejeon 305-348, Republic of Korea;
            mjkang@kasi.re.kr \\
       $^2$ Department of Astronomy and Space Science, 
            Kyung Hee University, Yongin-shi, Kyungki-do 446-701,
            Republic of Korea \\
       $^3$ Kapteyn Astronomical Institute, University of Groningen, 
            PO Box 800, 9700 AV, Groningen, The Netherlands \\
       $^4$ SRON Netherlands Institute for Space Research, 
            PO Box 800, 9700 AV, Groningen, The Netherlands \\
       $^5$ National Research Council of Canada,
            Herzberg Institute of Astrophysics, 5071 West Saanich Rd.,
            Victoria, BC, V9E 2E7, Canada \\
       $^6$ University of Victoria, Department of Physics and Astronomy,
            PO Box 3055, STN CSC, Victoria, BC, V8W 3P6, Canada \\
       $^7$ Department of Physics and Astronomy, Seoul National University,
            1 Gwanak-ro, Gwanak-gu, Seoul 151-742, Republic of Korea}

\begin{abstract}
The results of a maser survey
toward ninety-nine protostars in the Orion molecular cloud complex
are presented.
The target sources are low-mass protostars
identified from infrared observations.
Single-dish observations were carried out
in the water maser line at 22 GHz
and the methanol class I maser lines at 44, 95, and 133 GHz.
Most of the detected sources were mapped to determine the source positions.
Five water maser sources were detected,
and they are excited by HH 1--2 VLA 3, HH 1--2 VLA 1, L1641N MM1/3,
NGC 2071 IRS 1/3, and an object in the OMC 3 region.
The water masers showed significant variability 
in intensity and velocity with time scales of a month or shorter.
Four methanol emission sources were detected,
and those in the OMC 2 FIR 3/4 and L1641N MM1/3 regions
are probably masers.
The methanol emission from the other two sources
in the NGC 2071 IRS 1--3 and V380 Ori NE regions
are probably thermal.
For the water masers,
the number of detections per protostar in the survey region is about 2\%,
which suggests that the water masers of low-mass protostars are rarely detectable.
The methanol class I maser of low-mass protostars
is an even rarer phenomenon, with a detection rate much smaller than 1\%.
\end{abstract}


\section{INTRODUCTION}

The Orion giant molecular cloud complex
is located at a distance of $\sim$420 pc from the Sun
and is one of the nearest active star-forming regions
\citep{Menten07, Kim08}.
The molecular complex
largely consists of the Orion A and B clouds.
Numerous observations have been carried out
to investigate the star-formation activities in the Orion molecular clouds.
For example, there are surveys of dense cores
in molecular lines and dust continuum emission
\citep{Lada91, Tatematsu98, Lis98}
and surveys of outflows in the molecular hydrogen and CO lines
\citep{Davis09, Takahashi08}.
Recently, thousands of young stellar objects (YSOs) were identified
in the Orion molecular clouds based on infrared observations
with the {\it Spitzer Space Telescope},
and nearly five hundred objects among them
are likely protostars \citep{Megeath12}.
To study $\sim$300 of the {\it Spitzer}-identified Orion protostars,
the {\it Herschel} Orion Protostar Survey (HOPS) project was conducted
with the {\it Herschel Space Observatory}
\citep{Fischer10,Stutz13}.

Maser emission is an important signpost of star-formation regions
in the early stages of evolution.
Observations of masers allow detailed studies
of the small-scale environments of deeply embedded YSOs.
Many \water\ maser observations of YSOs have shown
that \water\ masers are usually distributed
very close ($\lesssim$ 1000 AU) to the central objects,
are highly variable in both intensity and velocity
with time scales from hours to years,
and trace molecular outflows and protostellar disks
\citep{Genzel77,Elitzur89,Comoretto90,
Torrelles98,Seth02,Furuya05,Goddi05,Felli07,Caswell10}.
Several \methanol\ maser lines also have been detected
toward star-forming regions \citep{Valtts95,Kurtz04,Kalenskii10}.
The methanol maser lines are divided
into class I  (36, 44, 84, and 95 GHz lines etc.)
and class II (6.7, 12, and 157 GHz lines etc.) \citep{Menten91}.
\methanol\ class I masers are usually offset by 0.1--1 pc
from star-formation phenomena
(such as hot molecular cores, ultracompact H {\small II} regions, and other maser sources)
and well-correlated with molecular outflows
\citep{Plambeck90, Cragg92, Kurtz04, Cyganowski09}.
In general, \methanol\ class I lines show little flux variability
\citep{Kurtz04,Kalenskii10}.

To investigate the star-formation activities
in the early stages,
we carried out a maser survey of {\it Spitzer}-identified protostars
distributed over the Orion molecular cloud complex.
Out of the protostars listed in the HOPS catalogue,
we selected protostars showing line wings
in the CO $J = 2 \rightarrow 1$ line spectra
obtained with the Seoul Radio Astronomy Observatory 6 m telescope.
All HOPS sources were observed in CO with a 48\arcsec\ beam 
down to an rms noise of 0.15 K or smaller.
The source selection was not affected by previously known masers.

In this paper, we present the results of the survey
in the \water\ and \methanol\ maser lines
with the Korean Very Long Baseline Interferometry Network (KVN) antennas.
In Section 2, we describe the KVN observations.
In Section 3, we present the results of the survey.
In Section 4, we describe the detected sources in detail.
In Section 5, we discuss the implications of the survey.
A summary is given in Section 6.

\begin{deluxetable}{lccccc}
\tabletypesize{\small}
\tablecaption{Telescope Parameters}
\tablewidth{0pt}
\tablehead{
\colhead{Telescope} & \colhead{Observing Period}
& \colhead{Frequency\tablenotemark{a}} & \colhead{Beam\tablenotemark{b}}
& \colhead{$\eta_A$\tablenotemark{c}} & \colhead{$f$\tablenotemark{d}} \\
&& \colhead{(GHz)} &\colhead{(arcsec)} && \colhead{(Jy K$^{-1}$)}}
\startdata
KVN Yonsei & 2010 Mar -- 2010 Jun & \phn22 &    122 & 0.65 & 15.3 \\
           &                      & \phn44 & \phn64 & 0.67 & 14.8 \\
           & 2012 Jan -- 2012 May & \phn22 &    119 & 0.65 & 15.3 \\
           &                      & \phn44 & \phn62 & 0.63 & 15.8 \\
           &                      & \phn86 & \phn32 & 0.48 & 20.7 \\
           &                      &    129 & \phn23 & 0.30 & 33.1 \\
KVN Ulsan  & 2012 Jan -- 2012 May & \phn22 &    120 & 0.62 & 16.0 \\
           &                      & \phn44 & \phn62 & 0.62 & 16.0 \\
           &                      & \phn86 & \phn33 & 0.49 & 20.3 \\
           &                      &    129 & \phn23 & 0.32 & 38.8 \\
KVN Tamna  & 2011 Nov -- 2012 Mar & \phn22 &    122 & 0.66 & 15.1 \\
           &                      & \phn44 & \phn64 & 0.60 & 16.6 \\
           & 2012 Mar -- 2012 May & \phn22 &    123 & 0.59 & 16.8 \\
           &                      & \phn44 & \phn62 & 0.62 & 16.0 \\
           &                      & \phn86 & \phn32 & 0.52 & 19.1 \\
           &                      &    129 & \phn22 & 0.40 & 31.1 
\enddata
\tablenotetext{a}{Frequency where the beam size and efficiency were measured.}
\tablenotetext{b}{Full width at half-maximum (FWHM) of the main beam.}
\tablenotetext{c}{Aperture efficiency.}
\tablenotetext{d}{Scaling factor for converting the KVN raw data
                  to spectra in the flux density scale for the maser lines,
                  which includes the telescope sensitivity,
                  quantization correction factor (1.25),
                  and sideband separation efficiency
                  (0.8 for the KVN Ulsan/Tamna 129 GHz band and 1.0 otherwise).}
\label{table1}
\end{deluxetable}

\section{OBSERVATIONS}

Ninety-nine protostars in the Orion molecular cloud complex were observed
using the KVN 21 m radio antennas in the single-dish telescope mode
during the 2010 and 2011--2012 observing seasons.
The observations were carried out
with the KVN Yonsei telescope at Seoul,
the KVN Ulsan telescope at Ulsan,
and the KVN Tamna telescope at Seogwipo, Korea.
The KVN telescopes are equipped with multi-frequency receiving systems
simultaneously operating at 22, 44, 86, and 129 GHz bands \citep{Han13}.
Telescope pointing was checked by observing Orion IRc2 \citep{Baudry95} 
in the SiO $v=1$ $J=1\rightarrow0$ maser line at 43 GHz. 
The pointing observations were performed about once every two hours. 
The rms pointing accuracy was better than $\sim$5\arcsec.
The alignments among the beams of different frequency bands are 
better than $\sim$3\arcsec\ \citep{Han13}. 

The target lines were
the \water\ $6_{16}\rightarrow5_{23}$ (22.23508 GHz) line
and the \methanol\ $7_{0} \rightarrow 6_{1}$ $A^{+}$,
$8_{0}\rightarrow 7_{1}$ $A^{+}$, and $6_{-1}\rightarrow5_{0}$ $E$
(44.06943, 95.169516, and 132.890800 GHz, respectively) lines.
The 4096-channel digital spectrometers were used as back ends.
A bandwidth of 32 MHz was selected,
which provides velocity resolutions of 0.105, 0.053, 0.0246, and 0.0176 \kms\
for the target lines.
For the 22 GHz \water\ and 44 GHz \methanol\ line spectra,
the Hanning smoothing was applied once and twice, respectively,
resulting in a velocity-channel width of 0.21 \kms\ in both lines.
For the 95 and 133 GHz \methanol\ line spectra,
the Hanning smoothing was applied three times,
which gives a velocity-channel width of 0.19 and 0.14 \kms, respectively.
For each observing season,
the data were calibrated using the standard efficiencies of KVN
listed in Table \ref{table1}
\citep{Lee11,Choi12} (http://kvn-web.kasi.re.kr).
The data were processed with the GILDAS/CLASS software
from Institut de RadioAstronomie Millim\'etrique
(http://www.iram.fr/IRAMFR/GILDAS).

In the 2010 season,
only the 22 GHz and 44 GHz band receivers were available,
and the \water\ line and the \methanol\ 44 GHz line were observed.
Integrations toward each target source were carried out
until the average spectra had a noise rms level down to $\sim$0.5 Jy.
System temperatures were
in the range of 70--190 K at 22 GHz and 140--270 K at 44 GHz.
Tables 2--4 list the target sources, coordinates, detectability,
and noise rms levels of the resulting spectra.
Table 2 lists the sources detected in either line.
When the emission from a single source is detected
toward several nearby target positions,
Table 2 lists only the position of the strongest signal.
Tables 3 and 4 list the sources undetected in either line.
The \water\ spectra of the sources marked by ``S'' in Tables 2 and 3
were affected by the Orion KL masers,
and it was difficult to separate the emission of the target region, if any,
from that of the Orion KL region (see below).
\begin{deluxetable}{lrrcccc}
\tabletypesize{\small}
\tablecaption{Target Sources with Detections}
\tablewidth{0pt}
\tablehead{
\colhead{HOPS\tablenotemark{a}} & \colhead{R.A.}  & \colhead{Decl.}
& \multicolumn{2}{c}{Detection\tablenotemark{b}}
& \colhead{$\sigma_{22}$} & \colhead{$\sigma_{44}$} \\
\cline{4-5}
& \colhead{(J2000.0)} & \colhead{(J2000.0)}
& \colhead{\water} & \colhead{\methanol} & \colhead{(Jy)} & \colhead{(Jy)}}
\startdata
 64 & 05 35 27.00 & $-$05 09 54.1 & S    & Y & 0.6 & 0.7 \\
 96 & 05 35 29.72 & $-$04 58 48.8 & S, Y & N & 0.6 & 0.5 \\
167 & 05 36 19.79 & $-$06 46 00.9 & Y    & N & 0.6 & 0.5 \\
182 & 05 36 18.83 & $-$06 22 10.2 & Y    & Y & 0.7 & 0.5 \\
361 & 05 47 05.16 &    00 21 35.9 & Y    & Y & 0.7 & 0.6 \\
362 & 05 36 36.10 & $-$06 38 53.9 & N    & Y & 0.5 & 0.5 \\
\enddata
\tablecomments{Units of Right Ascension are hours, minutes, and seconds,
               and units of declination are degrees, arcminutes,
               and arcseconds.}
\tablenotetext{a}{Source number in the HOPS catalog.}
\tablenotetext{b}{Y: Detection, N: Non-detection,
                  S: The \water\ spectrum is affected
                     by maser emission from the Orion KL region
                     coming through the beam side lobe.}
\label{table2}
\end{deluxetable}

\begin{deluxetable}{lrrcccc}
\tabletypesize{\footnotesize}
\tablecaption{Target Sources Affected by the Orion KL Water Masers}
\tablewidth{0pt}
\tablehead{
\colhead{HOPS} & \colhead{R.A.} & \colhead{Decl.}
& \multicolumn{2}{c}{Detection}
& \colhead{$\sigma_{22}$} & \colhead{$\sigma_{44}$} \\
\cline{4-5}
& \colhead{(J2000.0)} & \colhead{(J2000.0)}
& \colhead{\water} & \colhead{\methanol} & \colhead{(Jy)} & \colhead{(Jy)}}
\startdata
 10 & 05 35 09.01 & $-$05 58 27.6 & S & N & 0.6 & 0.5 \\
 11 & 05 35 13.41 & $-$05 57 58.1 & S & N & 0.4 & 0.5 \\
 12 & 05 35 08.60 & $-$05 55 54.3 & S & N & 0.5 & 0.6 \\
 17 & 05 35 07.18 & $-$05 52 05.9 & S & N & 0.5 & 0.6 \\
 19 & 05 35 25.99 & $-$05 51 22.9 & S & N & 0.4 & 0.6 \\
 20 & 05 33 30.71 & $-$05 50 41.0 & S & N & 0.6 & 0.6 \\
 41 & 05 34 29.44 & $-$05 35 42.7 & S & N & 0.5 & 0.5 \\
 44 & 05 35 10.57 & $-$05 35 06.3 & S & N & 0.5 & 0.6 \\
 48 & 05 35 06.56 & $-$05 32 51.6 & S & N & 0.5 & 0.5 \\
 49 & 05 34 48.88 & $-$05 31 45.9 & S & N & 0.7 & 0.7 \\
 50 & 05 34 40.91 & $-$05 31 44.4 & S & N & 0.5 & 0.6 \\
 51 & 05 35 15.83 & $-$05 30 05.5 & S & N & 0.5 & 0.6 \\
 53 & 05 33 57.37 & $-$05 23 30.4 & S & N & 0.5 & 0.5 \\
 55 & 05 33 54.09 & $-$05 21 49.5 & S & N & 0.5 & 0.6 \\
 57 & 05 35 19.84 & $-$05 15 08.5 & S & N & 0.6 & 0.6 \\
 58 & 05 35 18.51 & $-$05 13 38.2 & S & N & 0.5 & 0.7 \\
 60 & 05 35 23.33 & $-$05 12 03.1 & S & N & 0.5 & 0.5 \\
 61 & 05 33 25.91 & $-$05 12 02.6 & S & N & 0.5 & 0.6 \\
 65 & 05 35 21.56 & $-$05 09 38.7 & S & N & 0.7 & 0.4 \\
 70 & 05 35 22.41 & $-$05 08 04.8 & S & N & 0.6 & 0.7 \\
 73 & 05 35 27.70 & $-$05 07 03.5 & S & N & 0.5 & 0.6 \\
 75 & 05 35 26.66 & $-$05 06 10.3 & S & N & 0.7 & 0.5 \\
 77 & 05 35 31.53 & $-$05 05 47.3 & S & N & 0.6 & 0.6 \\
 80 & 05 35 25.19 & $-$05 05 09.5 & S & N & 0.5 & 0.6 \\
 82 & 05 35 19.73 & $-$05 04 54.6 & S & N & 0.4 & 0.6 \\
 86 & 05 35 23.65 & $-$05 01 40.3 & S & N & 0.6 & 0.6 \\
 89 & 05 35 19.96 & $-$05 01 02.6 & S & N & 0.5 & 0.6 \\
 90 & 05 35 34.47 & $-$05 00 52.0 & S & N & 0.7 & 0.6 \\
 93 & 05 35 15.03 & $-$05 00 08.2 & S & N & 0.5 & 0.6 \\
 95 & 05 35 34.20 & $-$04 59 52.2 & S & N & 0.4 & 0.5 \\
 97 & 05 35 28.89 & $-$04 57 38.9 & S & N & 0.6 & 0.6 \\
 98 & 05 35 19.32 & $-$04 55 44.9 & S & N & 0.6 & 0.5 \\
101 & 05 35 08.23 & $-$04 54 09.7 & S & N & 0.4 & 0.6 \\
349 & 05 35 26.21 & $-$05 08 33.0 & S & N & 0.7 & 0.5 \\
350 & 05 35 30.22 & $-$05 08 18.6 & S & N & 0.4 & 0.6 \\
351 & 05 35 31.43 & $-$05 04 46.7 & S & N & 0.6 & 0.5 \\
352 & 05 35 26.58 & $-$05 04 02.5 & S & N & 0.5 & 0.6 \\
\enddata
\tablecomments{See the end notes of Table 2.
               Sources listed in Table 2 are not repeated here.}
\label{table3}
\end{deluxetable}

\begin{deluxetable}{lrrccclrrcc}
\tabletypesize{\small}
\tablecaption{Target Sources without a Detection}
\tablewidth{0pt}
\tablehead{
\colhead{HOPS} & \colhead{R.A.} & \colhead{Decl.}
& \colhead{$\sigma_{22}$} & \colhead{$\sigma_{44}$}
&& \colhead{HOPS} & \colhead{R.A.} & \colhead{Decl.}
& \colhead{$\sigma_{22}$} & \colhead{$\sigma_{44}$} \\
& \colhead{(J2000.0)}& \colhead{(J2000.0)} & \colhead{(Jy)} & \colhead{(Jy)}
&&& \colhead{(J2000.0)}& \colhead{(J2000.0)} & \colhead{(Jy)} & \colhead{(Jy)}}
\startdata
  1 & 05 54 12.34 &    01 42 35.5 & 0.4 & 0.6 && 301 & 05 41 44.77 & $-$02 15 55.3 & 0.5 & 0.6 \\
  2 & 05 54 09.13 &    01 42 52.0 & 0.4 & 0.5 && 302 & 05 40 22.41 & $-$02 15 39.7 & 0.5 & 0.6 \\
124 & 05 39 19.98 & $-$07 26 11.2 & 0.5 & 0.6 && 303 & 05 42 02.62 & $-$02 07 45.7 & 0.5 & 0.6 \\
142 & 05 38 47.77 & $-$07 00 26.9 & 0.5 & 0.6 && 304 & 05 41 45.94 & $-$01 56 26.1 & 0.7 & 0.5 \\
152 & 05 37 58.76 & $-$07 07 25.3 & 0.5 & 0.5 && 305 & 05 41 45.38 & $-$01 51 56.8 & 0.6 & 0.6 \\
158 & 05 37 24.46 & $-$06 58 32.8 & 0.6 & 0.6 && 309 & 05 42 47.36 & $-$01 24 47.0 & 0.6 & 0.7 \\
164 & 05 37 00.45 & $-$06 37 10.5 & 0.4 & 0.6 && 310 & 05 42 27.68 & $-$01 20 01.0 & 0.5 & 0.6 \\
165 & 05 36 23.54 & $-$06 46 14.6 & 0.7 & 0.5 && 313 & 05 41 00.76 & $-$01 09 10.6 & 0.5 & 0.6 \\
166 & 05 36 25.13 & $-$06 44 41.8 & 0.7 & 0.5 && 316 & 05 46 07.29 &    00 13 23.0 & 0.6 & 0.6 \\
178 & 05 36 24.61 & $-$06 22 41.3 & 0.7 & 0.5 && 320 & 05 46 14.21 &    00 05 26.8 & 0.5 & 0.6 \\
180 & 05 36 59.39 & $-$06 10 15.6 & 0.5 & 0.6 && 321 & 05 46 33.17 &    00 00 02.2 & 0.5 & 0.6 \\
187 & 05 35 50.94 & $-$06 22 43.5 & 0.4 & 0.5 && 322 & 05 46 46.49 &    00 00 16.1 & 0.5 & 0.6 \\
191 & 05 36 17.26 & $-$06 11 11.0 & 0.6 & 0.5 && 325 & 05 46 39.25 &    00 01 15.0 & 0.6 & 0.7 \\
192 & 05 36 32.45 & $-$06 01 16.2 & 0.5 & 0.6 && 326 & 05 46 39.58 &    00 04 16.6 & 0.4 & 0.5 \\
221 & 05 42 47.05 & $-$08 17 07.0 & 0.6 & 0.5 && 330 & 05 46 51.37 &    00 19 47.4 & 0.4 & 0.5 \\
223 & 05 42 48.46 & $-$08 16 34.5 & 0.6 & 0.5 && 331 & 05 46 28.32 &    00 19 49.4 & 0.6 & 0.6 \\
256 & 05 40 45.26 & $-$08 06 42.2 & 0.5 & 0.6 && 332 & 05 47 31.70 &    00 20 20.8 & 0.3 & 0.4 \\
257 & 05 41 19.87 & $-$07 55 46.6 & 0.5 & 0.5 && 333 & 05 47 22.88 &    00 20 58.3 & 0.6 & 0.6 \\
260 & 05 40 19.39 & $-$08 14 16.4 & 0.5 & 0.6 && 334 & 05 46 48.52 &    00 21 28.2 & 0.7 & 0.6 \\
285 & 05 40 05.90 & $-$07 29 32.9 & 0.5 & 0.6 && 335 & 05 47 05.86 &    00 22 38.9 & 0.5 & 0.6 \\
287 & 05 40 08.78 & $-$07 27 27.7 & 0.4 & 0.5 && 336 & 05 46 02.28 &    00 23 30.7 & 0.5 & 0.6 \\
291 & 05 39 57.97 & $-$07 28 57.5 & 0.6 & 0.5 && 337 & 05 46 55.10 &    00 23 34.6 & 0.6 & 0.5 \\
292 & 05 37 54.88 & $-$07 41 20.3 & 0.5 & 0.6 && 340 & 05 47 01.29 &    00 26 21.5 & 0.4 & 0.6 \\
295 & 05 41 28.94 & $-$02 23 19.4 & 0.5 & 0.6 && 345 & 05 47 38.98 &    00 38 36.3 & 0.4 & 0.5 \\
296 & 05 41 17.17 & $-$02 18 07.6 & 0.6 & 0.6 && 347 & 05 47 15.89 &    00 21 23.8 & 0.7 & 0.7 \\
297 & 05 41 23.27 & $-$02 17 35.8 & 0.7 & 0.6 && 348 & 05 47 00.27 &    00 20 37.5 & 0.7 & 0.5 \\
298 & 05 41 37.17 & $-$02 17 17.0 & 0.6 & 0.5 && 354 & 05 54 24.26 &    01 44 19.3 & 0.5 & 0.6 \\
300 & 05 41 24.21 & $-$02 16 06.4 & 0.6 & 0.6 && 360 & 05 47 27.07 &    00 20 33.1 & 0.7 & 0.5 \\
\enddata
\tablecomments{See the end notes of Table 2.}
\label{table4}
\end{deluxetable}

In 2011 November--December,
all the target sources were observed again
in the \water\ line and the \methanol\ 44 GHz line.
Though the integrations were shorter in duration than those in the previous season,
close attention was paid to the effect of the Orion KL masers.
Observations toward the Orion KL region were made several times a day
so that the line profiles could be compared at any observing day.
The coordinates of Orion KL masers
used for this monitoring observations are 
(05$^{\rm h}$35$^{\rm m}$14\fs12, $-$05\arcdeg22$'$36\farcs4).
While the \water\ spectra of most of the ``S'' sources
showed only the signal from the Orion KL region,
one source (HOPS 96) showed a positive signal from the target source.

In 2012 January--May, the 86 and 129 GHz band receivers became available,
and we focused on the sources detected in the previous observing runs.
Observations were made in the \methanol\ 95 and 133 GHz lines
toward the sources detected in the 44 GHz line.
The areas around the detected sources were mapped
in the \water\ line and the \methanol\ 95 GHz line
to determine the source positions accurately
and to identify the YSOs associated with the emission sources.
The maps were made with grid spacings smaller than a half beam (FWHM/2). 

\subsection{Water Masers of the Orion KL Region}

As the KVN telescopes were designed mainly for interferometry,
the power levels of their beam side lobes are relatively high
\citep{Lee11}.
The Orion KL region contains a dense cluster of YSOs
generating bright \water\ masers \citep{Gaume98}. 
Our results show that the spectra toward the target sources
within $\sim$0.5\arcdeg\ from Orion KL
are contaminated by the emission from the Orion KL masers
coming through the side lobes.
The extent of the effect, however, varies
with the intensity of the Orion KL masers and the complex pattern
of the side lobes.
To verify if there is maser emission coming from a target source,
the spectra of the target source and Orion KL should be compared.

\begin{figure}[!t]
\epsscale{0.86}
\plotone{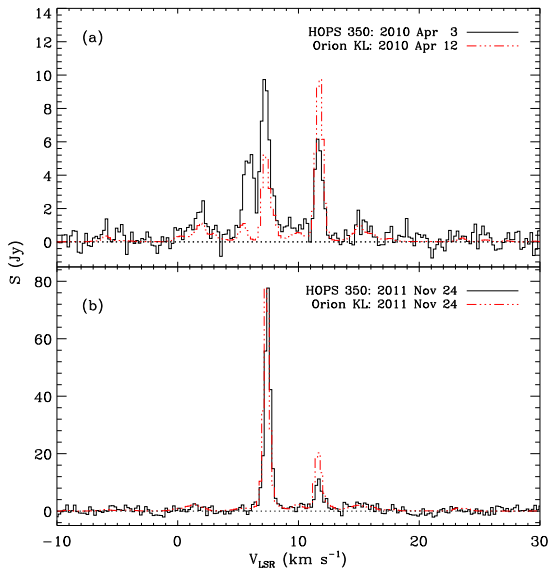}
\caption{
Spectra of the \water\ maser line
toward HOPS 350 (black solid line) and Orion KL (red dash-dotted line).
(a)
The spectrum toward HOPS 350 observed on 2010 April 3
and the spectrum toward Orion KL observed on 2010 April 12.
(b)
The maser spectra on 2011 November 24.
The Orion KL spectra are scaled down by a factor of 2200  in (a) and 800 in (b) for comparison.
}
\label{fig1}
\end{figure}

In the 2010 observing season,
the spectra of Orion KL were obtained sparsely,
and the comparison spectra are not available for every observing day,
which makes the comparison somewhat ambiguous.
Figure \ref{fig1}(a) shows an example.
HOPS 350 is located $\sim$0.25\arcdeg\ away from Orion KL.
All the velocity components in the spectrum of HOPS 350
can be seen in the spectrum of Orion KL obtained 9 days later.
However, the line profile (intensity ratios among the velocity components)
changed substantially,
and it is not clear
if any of the velocity components (e.g., the 6 \kms\ component)
contains emission from HOPS 350.
Such comparisons suggest
that the Orion KL masers should have been monitored daily.

In the 2011--2012 season,
while the target sources were observed again,
the spectra of Orion KL were obtained every $\sim$2 hours.
Figure \ref{fig1}(b) shows the spectra of the \water\ maser lines
toward HOPS 350 and Orion KL, observed on the same day.
The comparison shows
that all the velocity components came from the Orion KL masers.
The line profiles of the two spectra are slightly different
because the Orion KL masers are distributed
over an extended ($\sim$30$''$) region \citep{Gaume98}.

This contamination from Orion KL masers hinders
the detection of weak ($\lesssim$10 Jy) maser emission
from the affected target sources.
It is possible, however, to detect a maser towards these targets 
if the intensity is very strong
or if the line velocity is
far from the velocity interval crowded with the Orion KL masers.

\begin{deluxetable}{lcr@{ $\pm$ }lr@{ $\pm$ }lr@{ $\pm$ }lr}
\tabletypesize{\small}
\tablecaption{\water\ Maser Line Parameters}
\tablewidth{0pt}
\tablehead{
\colhead{HOPS} & \colhead{Epoch} & \multicolumn{2}{c}{$v_p$\tablenotemark{a}}
& \multicolumn{2}{c}{$\int F_{\nu} dv$}
& \multicolumn{2}{c}{$\delta{v}$\tablenotemark{a}} & \colhead{$F_p$} \\
&& \multicolumn{2}{c}{(\kms)} & \multicolumn{2}{c}{(Jy \kms)}
& \multicolumn{2}{c}{(\kms)} & \colhead{(Jy)}}
\startdata
 96 & 2011 Nov 24 & $-$14.39 & 0.06 &   2.4 & 0.5 & 0.55 & 0.14 &   4.1 \\
167 & 2010 Apr 20 &     6.69 & 0.01 &  37.9 & 0.4 & 0.68 & 0.01 &  52.8 \\
    & 2011 Nov 25 &  $-$0.08 & 0.01 &  31.4 & 0.9 & 0.92 & 0.02 &  26.8 \\
    &             &     0.91 & 0.01 &  79.4 & 1.3 & 0.92 & 0.02 &  81.1 \\
    & 2012 Jan 11 &     1.97 & 0.12 &   5.9 & 1.4 & 0.94 & 0.25 &   5.9 \\
    & 2012 Jan 22\tablenotemark{b}
                  &    13.42 & 0.03 &   8.0 & 0.9 & 0.59 & 0.07 &  12.9 \\
    & 2012 Jan 29 &  $-$0.14 & 0.05 &   1.5 & 0.2 & 0.67 & 0.11 &   2.2 \\
182 & 2010 Apr 24 &  $-$7.34 & 0.06 &   0.7 & 0.3 & 0.39 & 0.18 &   1.8 \\
    &             &     7.09 & 0.02 &  25.5 & 1.2 & 0.95 & 0.03 &  25.4 \\
    &             &     8.26 & 0.03 &  26.7 & 1.3 & 1.25 & 0.06 &  20.1 \\
    &             &    23.81 & 0.02 &   6.9 & 0.3 & 0.74 & 0.04 &   8.8 \\
    &             &    25.64 & 0.03 &   0.6 & 0.2 & 0.21 & 0.38 &   2.7 \\
    & 2011 Nov 25 &     6.62 & 0.01 & 240.4 & 0.7 & 0.97 & 0.01 & 233.7 \\
    & 2012 Jan 11 &     6.45 & 0.01 & 296.4 & 0.8 & 0.84 & 0.01 & 331.3 \\
    & 2012 Jan 30 &     6.40 & 0.01 & 239.4 & 0.4 & 0.78 & 0.01 & 288.2 \\
361 & 2010 May 26 &     7.54 & 0.01 & 129.5 & 0.7 & 0.79 & 0.01 & 155.0 \\
    &             &     8.98 & 0.01 &  90.2 & 0.8 & 1.16 & 0.01 &  73.4 \\
    & 2011 Nov 25 &  $-$2.74 & 0.03 &   4.0 & 0.7 & 0.45 & 0.09 &   8.5 \\
    &             &  $-$0.82 & 0.02 &  13.2 & 0.8 & 0.64 & 0.04 &  19.4 \\
    &             &     3.36 & 0.02 &  39.0 & 1.6 & 0.79 & 0.04 &  46.2 \\
    &             &     5.79 & 0.06 &  26.5 & 2.5 & 1.43 & 0.17 &  17.4 \\
    &             &     7.78 & 0.03 &  46.8 & 2.9 & 1.10 & 0.08 &  39.8 \\
    &             &     9.14 & 0.05 &  17.5 & 2.5 & 0.92 & 0.14 &  18.0 \\
    &             &    10.82 & 0.11 &  12.4 & 2.2 & 1.54 & 0.34 &   7.6 \\
    &             &    12.63 & 0.01 & 123.6 & 1.4 & 0.80 & 0.01 & 145.1 \\
    & 2012 Jan 11 &  $-$5.82 & 0.02 &  22.7 & 1.4 & 0.60 & 0.04 &  35.7 \\
    &             &     2.85 & 0.01 & 280.8 & 1.1 & 0.75 & 0.01 & 350.1 \\
    &             &     5.89 & 0.03 &  18.3 & 1.7 & 0.69 & 0.09 &  25.1 \\
    & 2012 Jan 30 &  $-$5.62 & 0.06 &   7.3 & 0.7 & 1.02 & 0.08 &   6.8 \\
    &             &     2.51 & 0.01 & 310.7 & 0.8 & 0.74 & 0.01 & 394.7 \\
    &             &     5.82 & 0.01 &  16.8 & 0.5 & 0.85 & 0.03 &  18.6 \\
    &             &     8.01 & 0.05 &  14.4 & 0.9 & 2.01 & 0.17 &   6.7 \\
    &             &    13.29 & 0.01 &  58.2 & 0.4 & 0.77 & 0.01 &  71.3 \\
    &             &    16.25 & 0.01 &  20.8 & 0.5 & 0.86 & 0.02 &  22.7 \\
\enddata
\tablecomments{The line parameters are from the Gaussian fits
               to each velocity component in the spectra
               observed toward the target source positions in Table 2.}
\tablenotetext{a}{When the uncertainty is smaller than 0.015 \kms,
                  it is listed as 0.01 \kms.}
\tablenotetext{b}{The parameters are for the spectrum
                  toward the KLC 3 position in Table \ref{table6}.}
\label{table5}
\end{deluxetable}

\section{RESULTS}

\subsection{\water\ Masers}

The \water\ maser line was detected toward four target sources:
HOPS 96, 167, 182, and 361.
All the \water\ masers showed large variations in flux and velocity
over the observing runs from 2010 to 2012.
Table \ref{table5} lists the properties of the detected \water\ masers:
peak velocity, integrated line flux, line FWHM, and peak intensity
from Gaussian fits.
While the HOPS 96 maser was detected only once,
the others were detected multiple times.
Even for the multiply detected target sources,
the line velocities of detected spectral features
changed significantly from one observing run to the next,
except for the steady $\sim$6.6 \kms\ component of HOPS 182.
This variability suggests 
that the typical lifetime of each velocity component
is about a month or shorter.

The areas around the detected target sources were mapped
to identify the YSOs responsible for the excitation of the masers.
The HOPS 96 field was not mapped
because the maser detected in 2011 November
already disappeared by the time we tried to map it in 2012 January.
Each map was fitted with a Gaussian intensity profile
having the same FWHM as the main beam.
The intensity distribution of each mapping field and each velocity component
is consistent with what is expected from a point-like source
(convolved with the beam).
Table \ref{table6} lists the source positions determined by the mapping.
Two \water\ maser sources (KLC 2/3) were identified in the HOPS 167 field.
For the HOPS 361 field,
all the velocity components (at the mapping epoch)
seem to come from a single source
(or a single region much smaller than the beam size), KLC 6.

\begin{deluxetable}{lccccrrc}
\tabletypesize{\small}
\tablecaption{Line Emission Source Positions Determined by Mapping}
\tablewidth{0pt}
\tablehead{
\colhead{KLC\tablenotemark{a}} & \colhead{HOPS} & \colhead{Line}
& \colhead{Epoch} & \colhead{$v_p$}
& \colhead{R.A.} & \colhead{Decl.} & \colhead{Associated Object} \\
& \colhead{Field} &&& \colhead{(\kms)} &
\colhead{(J2000.0)} & \colhead{(J2000.0)} &}
\startdata
1 &  64 & \methanol & 2012 May  7     &   11 & 05 35 27.7 & $-$05 09 46
        & OMC 2 MIR 23 \\
2 & 167 & \water    & 2011 Nov 25     & 0, 1 & 05 36 18.6 & $-$06 45 24
        & HH 1--2 VLA 3 \\
3 &     & \water    & 2012 Jan 22     &   13 & 05 36 22.5 & $-$06 46 01
        & HH 1--2 VLA 1 \\
4 & 182 & \water    & 2012 Jan 11     &    6 & 05 36 18.4 & $-$06 22 11
        & L1641N MM1/3 \\
5 &     & \methanol & 2012 May  7     &    7 & 05 36 17.7 & $-$06 22 20
        & L1641N MM1/3 \\
6 & 361 & \water    & 2011 Nov 25     & ($-$3, 13)\tablenotemark{b}
                                             & 05 47 04.2 &    00 21 44
        & NGC 2071 IRS 1/3 \\
7 &     & \methanol & 2012 Jan 30     &   10 & 05 47 04.9 &    00 21 44
        & NGC 2071 IRS 1--3 \\
8 & 362 & \methanol & 2012 May 11--12 &    9 & 05 36 36.7 & $-$06 39 17
        & V380 Ori NE \\
\enddata
\tablecomments{For most sources,
               the position uncertainty from the Gaussian fit to the map
               is much smaller than the pointing uncertainty of the telescopes,
               and their total position uncertainty is $\sim$5$''$.
               For KLC 2 and 3, the two contributions are comparable,
               and the total position uncertainty is $\sim$6$''$.
               The \methanol\ source positions were determined
               from the maps made in the 95 GHz line.}
\tablenotetext{a}{The line emission sources are numbered with a prefix KLC.}
\tablenotetext{b}{All the velocity components listed in Table 5.}
\label{table6}
\end{deluxetable}

\subsection{\methanol\ Lines}

The \methanol\ 44 GHz line was detected toward four target sources:
HOPS 64, 182, 361, and 362.
Follow-up observations toward these sources showed
that the 95 and 133 GHz lines are also detectable.
For each source, the three \methanol\ lines
have similar velocities and profiles,
which suggests that they have the same origin.
The 95 GHz line maps of HOPS 64/182 show
that these emission sources are compact
(much smaller than the $\sim$29$''$ beam size).
The maps of HOPS 361/362 show
that these sources are extended (20$''$--50$''$).
Table \ref{table6} lists the peak positions,
and Table \ref{table7} lists the properties of the \methanol\ lines
at those positions.

\begin{deluxetable}{lccr@{ $\pm$ }lr@{ $\pm$ }lr@{ $\pm$ }lr}
\tabletypesize{\small}
\tablecaption{\methanol\ Line Parameters}
\tablewidth{0pt}
\tablehead{
\colhead{KLC} & \colhead{HOPS} & \colhead{Frequency}
& \multicolumn{2}{c}{$v_p$\tablenotemark{a}}
& \multicolumn{2}{c}{$\int F_{\nu} dv$}
& \multicolumn{2}{c}{$\delta{v}$\tablenotemark{a}} & \colhead{$F_p$} \\
& \colhead{Field} & \colhead{(GHz)} & \multicolumn{2}{c}{(\kms)}
& \multicolumn{2}{c}{(Jy \kms)} & \multicolumn{2}{c}{(\kms)} & \colhead{(Jy)}}
\startdata
1 &  64 & \phn44 & 11.24 & 0.01 & 101.5 & 0.3 & 0.41 & 0.01 & 234.9 \\
  &     & \phn95 & 11.44 & 0.01 & 139.8 & 1.0 & 0.60 & 0.01 & 219.1 \\
  &     &    133 & 11.37 & 0.01 &  35.1 & 1.4 & 0.66 & 0.04 &  50.2 \\
5 & 182 & \phn44 &  7.35 & 0.03 &   6.1 & 0.2 & 1.80 & 0.07 &   3.2 \\
  &     & \phn44\tablenotemark{b}
                 &  7.00 & 0.21 &   3.3 & 0.2 & 1.02 & 0.21 &   3.0 \\
  &     &        &  7.90 & 0.21 &   1.9 & 0.2 & 0.60 & 0.21 &   3.0 \\
  &     & \phn95 &  7.42 & 0.04 &  14.0 & 0.7 & 1.63 & 0.09 &   8.1 \\
  &     &    133 &  7.30 & 0.18 &  12.0 & 1.8 & 2.67 & 0.53 &   4.2 \\
7 & 361 & \phn44 &  9.75 & 0.10 &   6.4 & 0.4 & 3.32 & 0.33 &   1.8 \\
  &     & \phn95 &  9.53 & 0.08 &  13.6 & 0.8 & 2.92 & 0.22 &   4.4 \\
  &     &    133 &  9.54 & 0.10 &  34.6 & 2.4 & 3.30 & 0.31 &   9.8 \\
8 & 362 & \phn44 &  8.48 & 0.05 &  12.1 & 0.4 & 3.26 & 0.15 &   3.5 \\
  &     & \phn95 &  8.57 & 0.07 &  26.3 & 1.1 & 3.58 & 0.18 &   6.9 \\
  &     &    133 &  8.80 & 0.22 &  49.6 & 3.5 & 6.25 & 0.59 &   7.4 \\
\enddata
\tablecomments{The line parameters are from the Gaussian fits
               to the spectra observed toward the source positions in Table
6.}
\tablenotetext{a}{When the uncertainty is smaller than 0.015 \kms,
                  it is listed as 0.01 \kms.}
\tablenotetext{b}{Two-component Gaussian fit.}
\label{table7}
\end{deluxetable}

In contrast with the \water\ masers,
the \methanol\ lines did not show any significant variability.
The flux variability, if any, is smaller
than the calibration uncertainty of the telescopes ($\sim$10\%).
The peak velocities of the \methanol\ lines
are always close (within $\sim$1 \kms)
to the systemic velocity of the ambient cloud.

Unlike with the \water\ maser line,
the detection of \methanol\ class I maser lines
does not necessarily mean that the detected flux is amplified emission.
They can be either maser or thermal emission, or a mixture of both.
The \methanol\ lines of KLC 1 are particularly strong and narrow,
and they are most likely masers.
The KLC 5 lines are probably partial masers
because the lines are narrow and the source size is small.
The KLC 7 and 8 lines are probably thermal
because the lines are relatively wide and the sources are extended.

\section{NOTES ON THE DETECTED SOURCES}

In this section, we describe the detected sources in detail
and present the spectra and maps of the \water\ and \methanol\ lines.
The maps show the source positions determined by mapping observations,
as well as the positions of known YSOs
superposed on the color-composite infrared images
of the Wide-Field Infrared Survey Explorer (WISE) \citep{Wright10}.

\subsection{KLC 1 in the HOPS 64 Field (OMC 2)}

\begin{figure}[!t]
\plotone{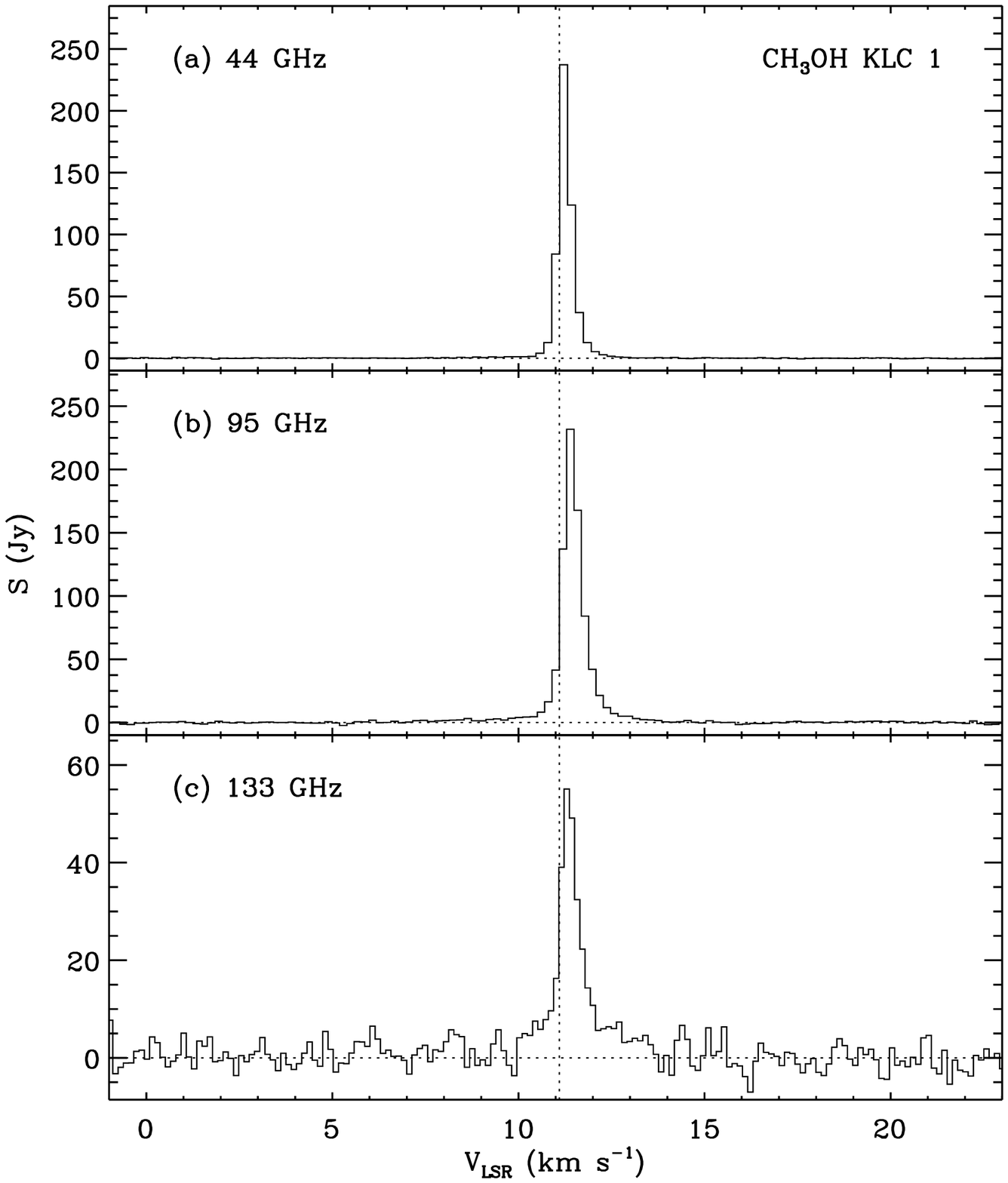}
\caption{
Spectra of the \methanol\ class I maser lines toward KLC 1.
(a) \mfour\ line. (b) \mnine\ line. (c) \mone\ line.
The vertical dotted line indicates
the systemic velocity of the ambient dense gas, $v_{\rm LSR}$ = 11.1 \kms\
\citep{Tatematsu98}.
}
\label{fig2}
\end{figure}

\begin{figure}[!t]
\plotone{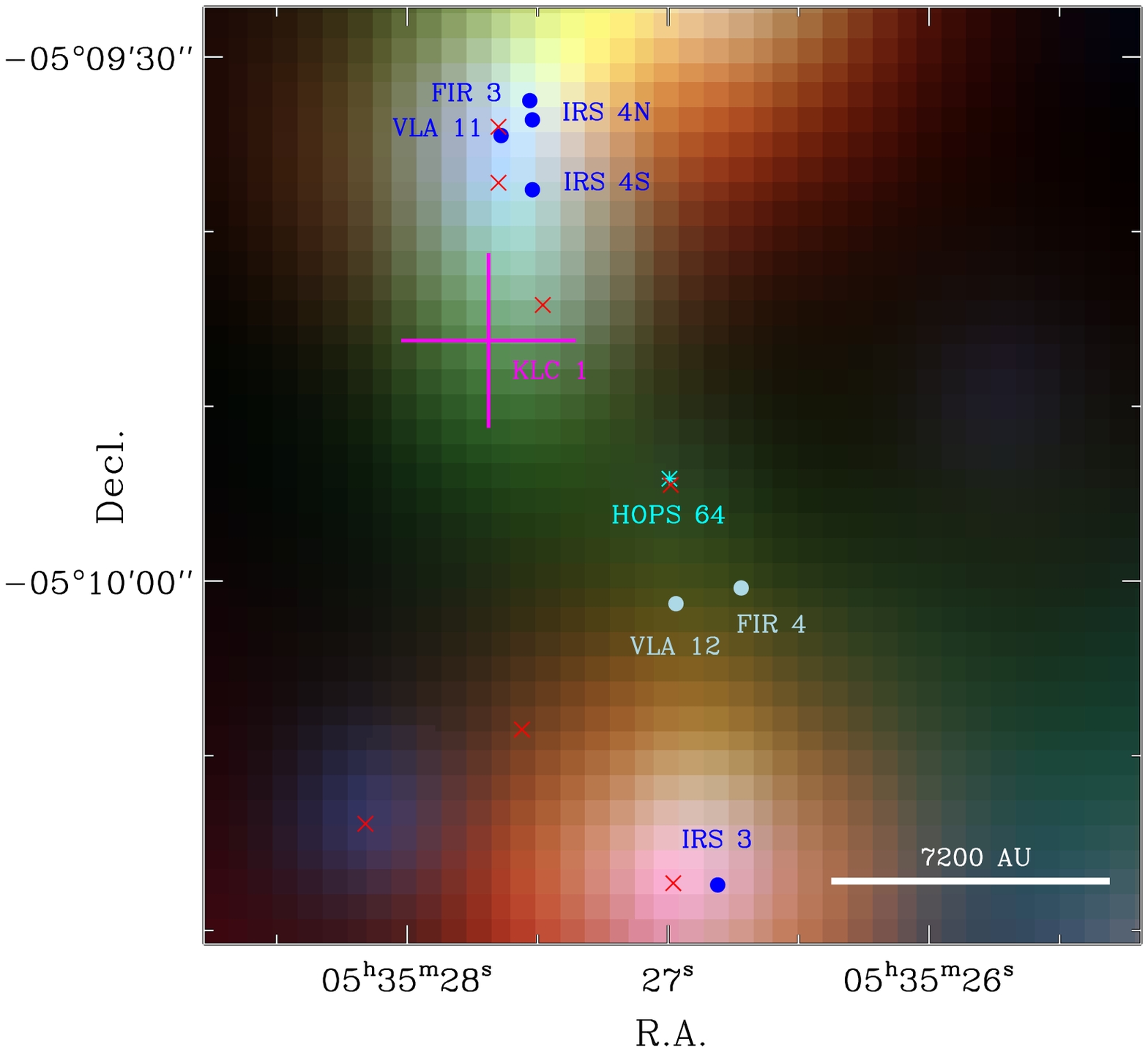}
\caption{
Position of KLC 1 (magenta plus sign) in the HOPS 64 (asterisk) field.
The size of the plus sign corresponds to the positional uncertainty.
The filled circles mark the positions of YSOs:
near-IR sources (IRS 3/4N/4S), 1.3 mm continuum sources (FIR 3/4),
and 3.6 cm continuum sources (VLA 11/12)
\citep{Chini97,Gatley74,Pendleton86,Reipurth99}.
The crosses mark the mid-IR sources MIR 21--27, from north to south
 \citep{Nielbock03}.
The scale bar in the right bottom corner corresponds
to 7200 AU at a distance of 420 pc (FWHM/2 of the 95 GHz main beam).
The background color image is composed of WISE 12 \micron\ (red),
4.6 \micron\ (green), and 3.4 \micron\ (blue) maps \citep{Wright10}. 
}
\label{fig3}
\end{figure}

The \methanol\ class I maser lines at 44, 95, and 133 GHz
were detected toward KLC 1 (Figure \ref{fig2}).
All the three lines are quite strong.
Their velocities are close (within 0.3 \kms)
to the systemic velocity of the ambient dense molecular cloud
measured in the CS $J = 2 \rightarrow 1$ line \citep{Tatematsu98}.
The source position was determined using a 95 GHz line map
that is a regular-grid map of 5 $\times$ 5 points
with a spacing of 7\farcs5.
Fitting the map with a Gaussian intensity profile suggests
that the source is compact (much smaller than the beam size).
The best-fit source position of KLC 1
is (10\arcsec, 8\arcsec) with respect to HOPS 64 (Figure \ref{fig3}).
KLC 1 is located between the FIR 3 cluster (FIR 3, IRS 4N/S, and VLA 11)
and the FIR 4 cluster (FIR 4, HOPS 64, and VLA 12),
and it is difficult to specify the YSO
responsible for the excitation of the maser.

KLC 1 positionally coincides with the mid-IR source MIR 23 within 4$''$
\citep{Nielbock03}.
There is little known about MIR 23,
and its relation with the maser activities is not clear.
Interferometric observations in the 44 and 95 GHz lines
were presented by \cite{Slysh09},
which show several maser spots distributed along a line.
Their maser spot A coincides with KLC 1 within 1$''$,
while other spots are also within the uncertainty circle of KLC 1.
\cite{Slysh09} suggested that the maser is associated with IRS 4S.
The linear distribution of the spots, however,
is neither pointing toward nor perpendicular to IRS 4S.
The line of spots rather points toward the FIR 4 cluster.
Further observations are needed to understand the KLC 1 maser
and its relation with the YSOs around it.

\begin{figure}[!t]
\plotone{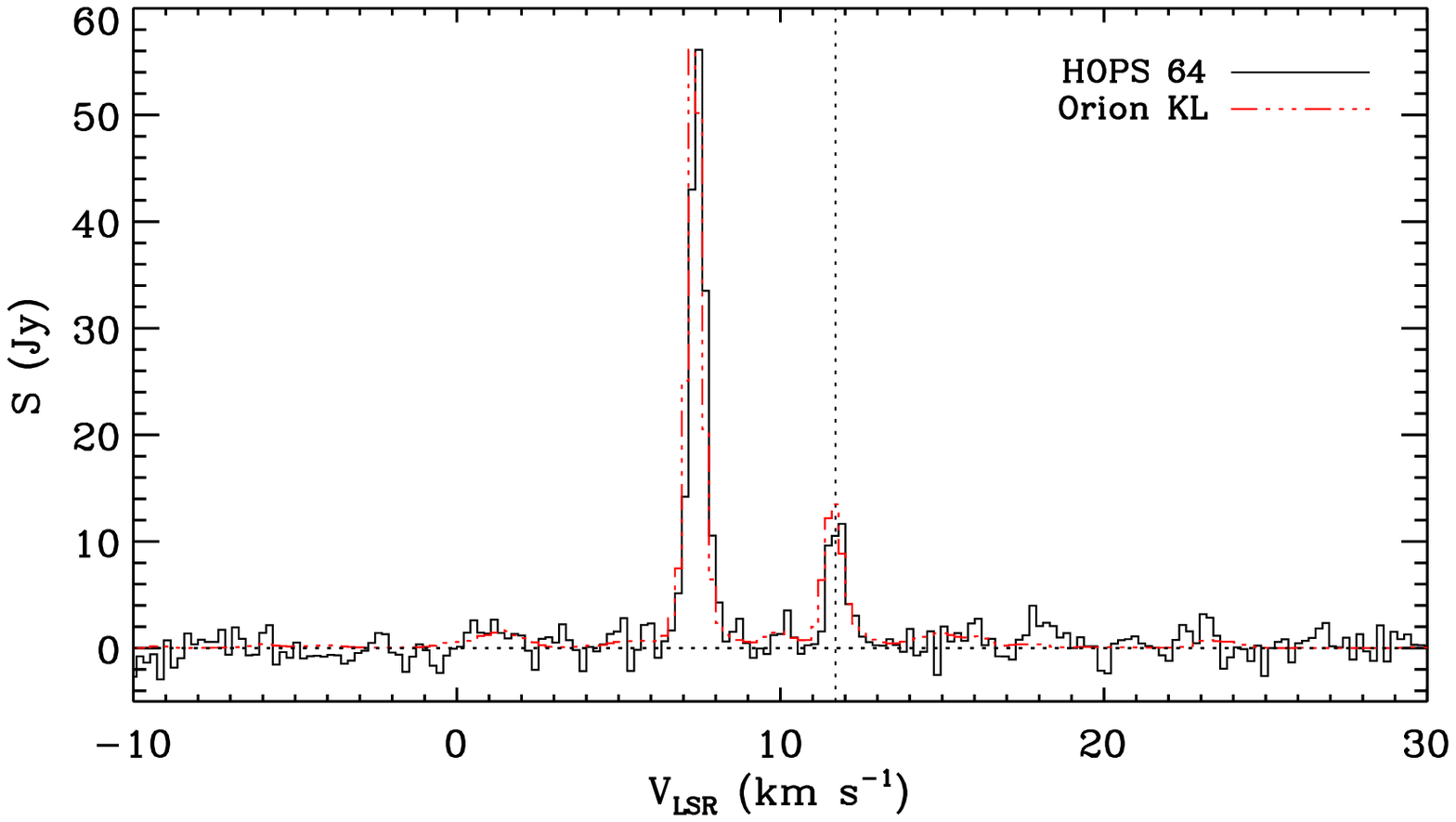}
\caption{
Spectra of the \water\ line
toward HOPS 64 (black) and Orion KL (red) on 2011 November 23.
The Orion KL spectrum is scaled down by a factor of 1300 for comparison.
The vertical dotted line indicates the systemic velocity.
}
\label{fig4}
\end{figure}

The \water\ line spectra toward HOPS 64 were affected by the Orion KL masers.
Figure \ref{fig4} shows an example.
All the velocity components in the HOPS 64 spectrum
may be attributed to the Orion KL masers,
and no \water\ maser associated specifically with HOPS 64 was detected in this survey.
Detections of an \water\ maser in this region, however, 
were reported previously \citep{Morris76,Genzel79}.
Interferometric observations showed
that the \water\ maser is associated with FIR 3 \citep{Tofani95}.
The \water\ maser was probably inactive in the 2010--2012 period.

\subsection{HOPS 96 (OMC 3)}
\label{sec:hops096}

There are at least three YSOs in the HOPS 96 field:
HOPS 96 in the SIMBA a condensation 
and MIR 1/2 in the SIMBA c condensation \citep{Nielbock03}.
These condensations show molecular outflows
traced by the CO $J = 3 \rightarrow 2$ line \citep{Takahashi08}.

\begin{figure}[!t]
\plotone{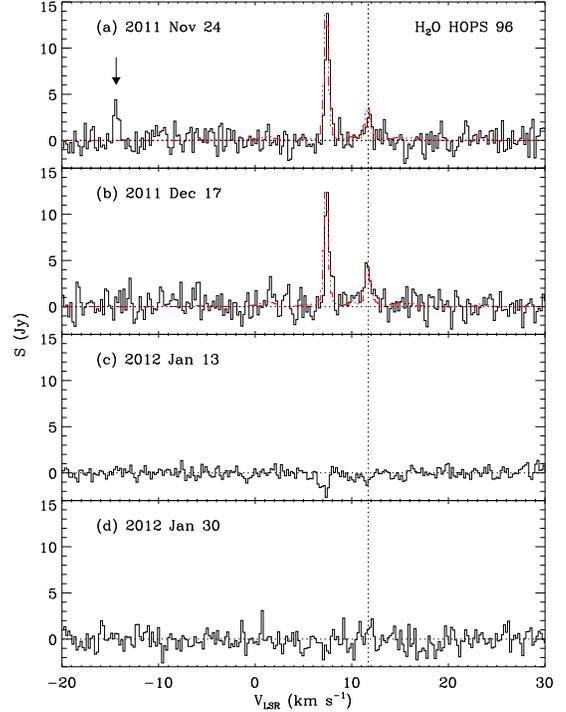}
\caption{
Spectra of the \water\ line toward HOPS 96  (black) and Orion KL (red).
(a)
The spectra observed on 2011 November 24.
The arrow marks the velocity component attributable to HOPS 96.
(b--d)
The spectra observed on 2011 December 17, 2012 January 13,
and 2012 January 30, respectively.
The Orion KL spectra are scaled down by a factor of 4400 in (a) and 3600 in (b) for comparison.
The vertical dotted line indicates
the systemic velocity of the ambient dense gas, $v_{\rm LSR}$ = 11.7 \kms\
\citep{Tatematsu93}.
}
\label{fig5}
\end{figure}

\begin{figure}[!p]
\plotone{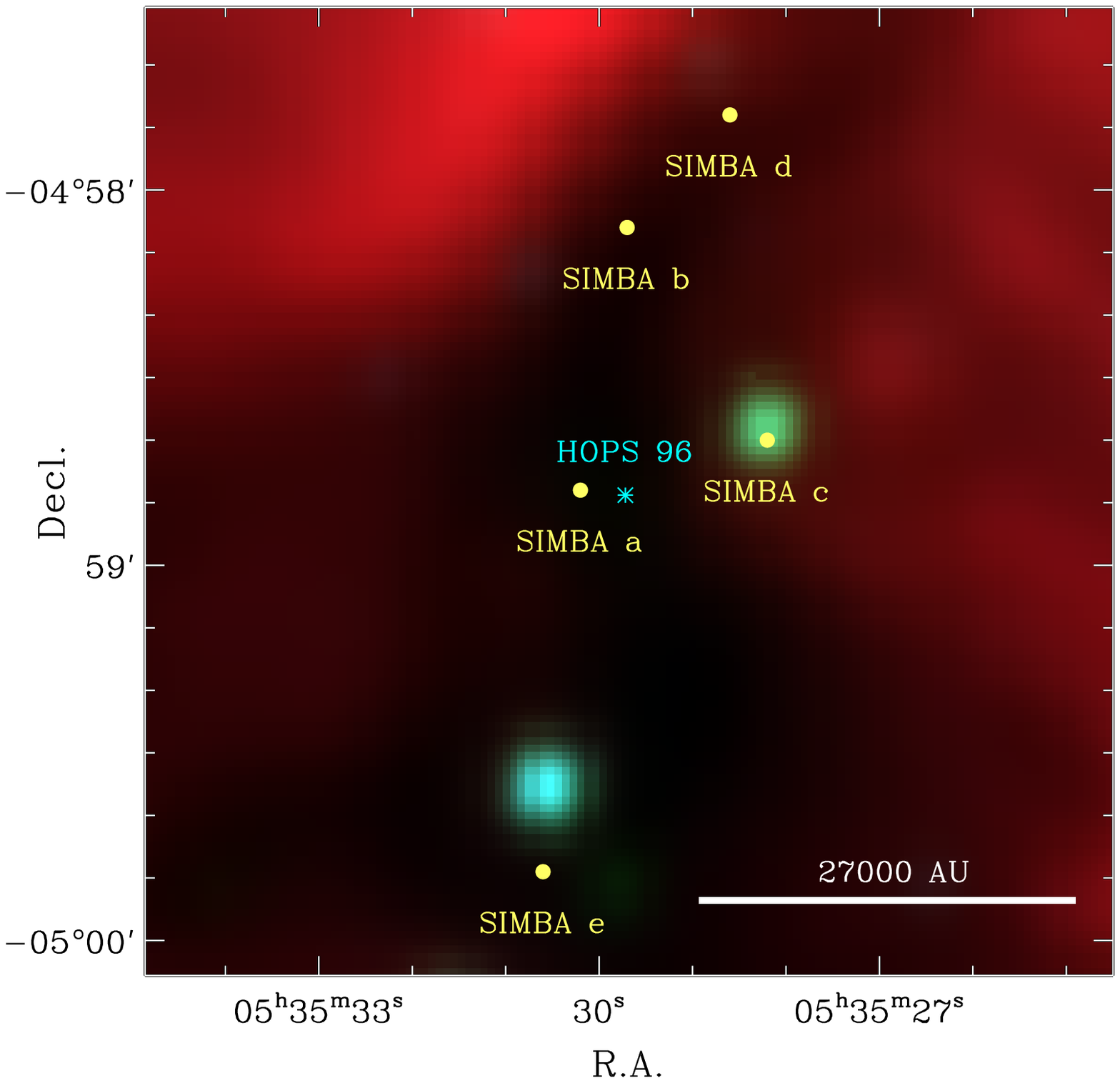}
\caption{
Positions of YSOs in the HOPS 96 (asterisk) field (OMC 3 SIMBA).
The filled circles mark the dense cores (condensations)
traced by the 1.2 mm continuum emission \citep{Nielbock03}.
There are two mid-IR sources associated with condensation c
\citep{Nielbock03}.
The scale bar in the right bottom corner corresponds
to 27000 AU at a distance of 420 pc (FWHM/2 of the 22 GHz main beam).
}
\label{fig6}
\end{figure}

Though the \water\ spectra toward HOPS 96 are
occasionally contaminated by the Orion masers,
an emission component was marginally detected in the uncontaminated velocity space
(Figure \ref{fig5}).
This velocity component detected at $-$14 \kms\ in 2011 November
may be coming from a maser source in the HOPS 96 field.
The component, however,
disappeared in all subsequent observing runs,
and there is no map to determine its position accurately.
The maser source can be anywhere in the region shown in Figure \ref{fig6} 
and may be excited by one of the three YSOs listed above.

\subsection{KLC 2/3 in the HOPS 167 Field (HH 1)}

HOPS 167 is located in the HH 1--2 region that is an extensively
studied star-forming location 
\citep{Pravdo85,Reipurth93,Choi97,Rodriguez00,Kwon10,Fischer10}.
There are two class 0 protostars (VLA 1/3) though the classification
is somewhat uncertain \citep{Chini01}. \cite{Fischer10} reported that both
VLA 1 (HOPS 203) and VLA 3 (HOPS 168) are in an active state of
mass infall and accretion.  They are probably the youngest and most
luminous protostars in this region.

\begin{figure}[!p]
\plotone{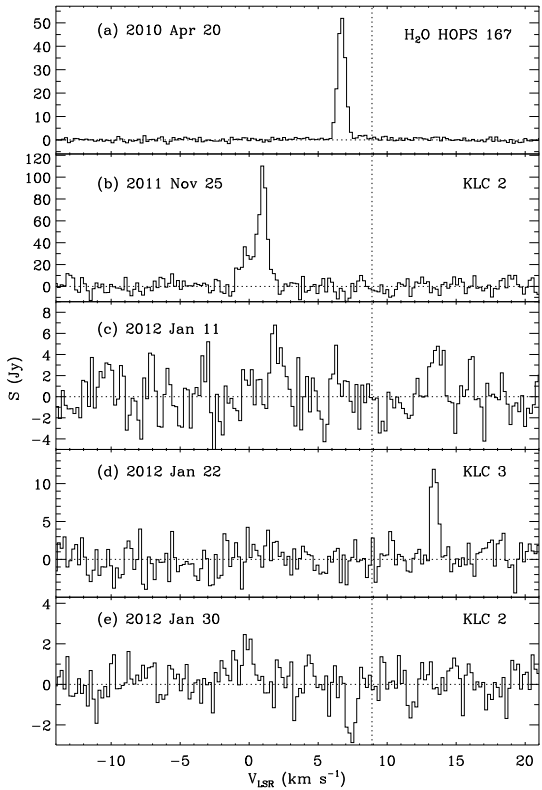}
\caption{
Spectra of the \water\ line.
The observing epochs are labeled in each panel.
(a, c)
Spectra toward HOPS 167.
(b, e)
Spectra toward KLC 2.
(d)
Spectrum toward KLC 3.
The vertical dotted line indicates
the systemic velocity of the ambient dense gas, $v_{\rm LSR}$ = 8.9 \kms\
\citep{Tatematsu98}.
}
\label{fig7}
\end{figure}

The HOPS 167 region displayed interesting \water\ maser activities.
The maser emission was relatively strong (30--80 Jy) in 2010--2011,
became weak (2--13 Jy) in 2012 January,
and disappeared in 2012 May (Figure \ref{fig7}).

\begin{figure}[!t]
\plotone{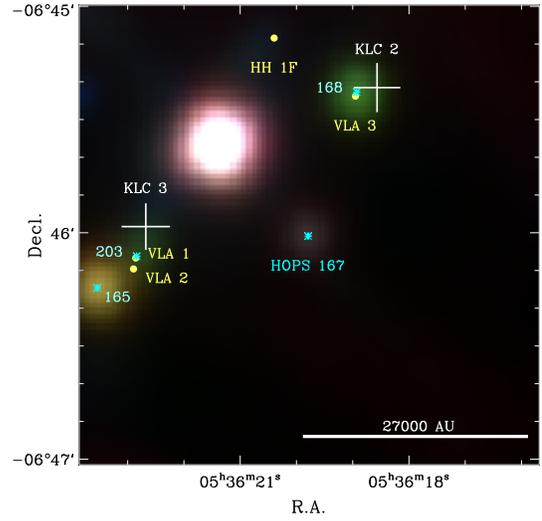}
\caption{
Positions of KLC 2/3 (white plus signs) in the HOPS 167 field.
The asterisks mark HOPS sources \citep{Fischer10}.
The filled circles mark the radio continuum sources \citep{Rodriguez90}.
}
\label{fig8}
\end{figure}

Mapping observations were carried out three times.
In 2011 November,
the HOPS 167 region was mapped with a grid spacing of 65\arcsec.
The best-fit source position of KLC 2
is ($-$18$''$, 37$''$) with respect to HOPS 167 (Figure \ref{fig8}).
On 2012 January 22, the region was mapped again with a 43$''$ spacing.
The position of the 13 \kms\ component
turned out to have an offset from KLC 2 by about a half beam,
and there was no emission from KLC 2.
The best-fit position of this new source, KLC 3,
is (40$''$, 0$''$) with respect to HOPS 167.
The separation between KLC 2 and 3 is $\sim$69$''$.
On 2012 January 29--30,
the region was mapped again with a 43$''$ spacing.
The position of the 0 \kms\ component
was consistent with KLC 2, though the detection was marginal,
and emission from KLC 3 had disappeared.
In short, two \water\ maser sources were detected in the HOPS 167 field,
and both of them displayed a rapid variability.

KLC 2 coincides within 5$''$
with the deeply embedded protostar VLA 3 \citep{Rodriguez00}.
The KLC 2 (VLA 3) maser was reported previously
and known to vary rapidly \citep{Lo75,Ho82,Haschick83}.
VLA 3 drives a molecular outflow that is separate from the HH 1--2 outflow
\citep{Choi97,Moro-Martin99}.

KLC 3 is located $\sim$8$''$ northwest of VLA 1 (HOPS 203),
which corresponds to the brightest knot of the HH 1 jet
(knot G in Figure 3 of \cite{Reipurth93}).
Considering the 5$''$ pointing uncertainty of the telescope,
KLC 3 may be associated with either the HH 1 jet or VLA 1.
In any case, VLA 1 seems to be the YSO
responsible for the excitation of the KLC 3 maser.
Detection of \water\ maser emission near VLA 1
has not been reported previously.

\subsection{KLC 4/5 in the HOPS 182 Field (L1641N)}

The central region of the L1641N cluster is crowded
with many YSOs and multiple outflows \citep{Stanke07,Galfalk08},
and it is difficult to study the nature of each individual object.
MM1 (HOPS 182) is a low-mass or intermediate-mass protostar
deeply embedded in an envelope of $\sim$1.6 \msun\ \citep{Chen95,Stanke07}.
MM3 is a ``protrusion'' of MM1
in the 1.3 mm continuum image of \cite{Stanke07}.
It is probably a deeply embedded YSO, but its nature is poorly known.

\begin{figure}[!t]
\plotone{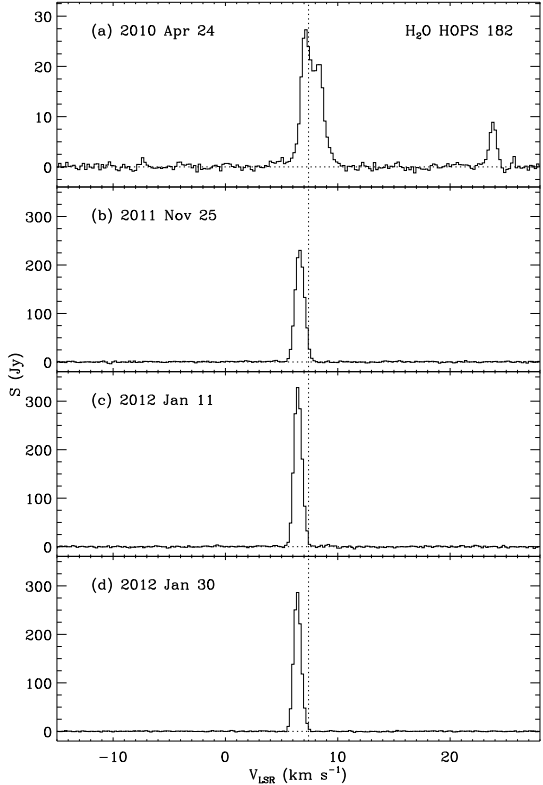}
\caption{
Spectra of the \water\ line toward HOPS 182 (KLC 4).
The vertical dotted line indicates
the systemic velocity of the ambient dense gas, $v_{\rm LSR}$ = 7.4 \kms\
\citep{Tatematsu98}.
}
\label{fig9}
\end{figure}

\begin{figure}[!t]
\plotone{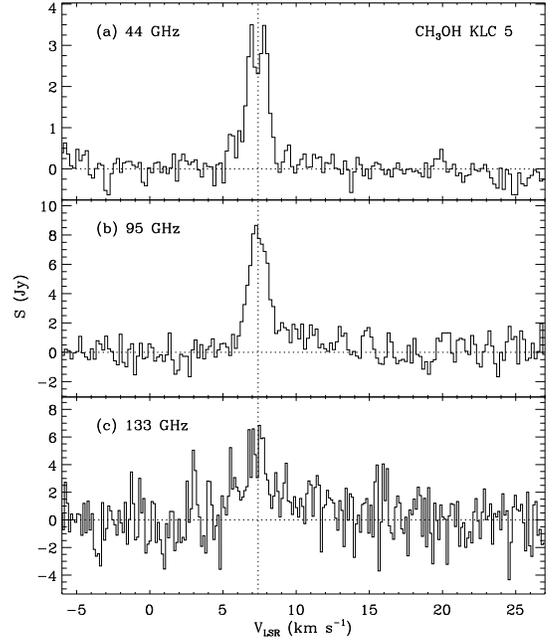}
\caption{
Spectra of the \methanol\ 44, 95, and 133 GHz lines toward KLC 5.
The vertical dotted line indicates the systemic velocity.
}
\label{fig10}
\end{figure}

The \water\ line and the three \methanol\ lines
were detected toward the L1641N region (Figures \ref{fig9}--\ref{fig10}).
The \water\ maser spectrum
displayed five velocity components in 2010 April:
two near the systemic velocity
and three at relatively high velocities
(one highly blueshifted and two highly redshifted).
In the subsequent observing runs,
the high-velocity components disappeared
while one of the low-velocity components became much brighter than before.

\begin{figure}[!b]
\plotone{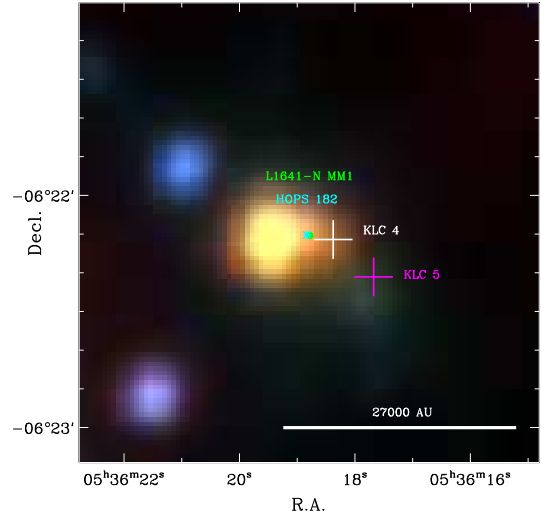}
\caption{
Positions of KLC 4/5 (white/magenta plus signs)
in the HOPS 182 (asterisk) field.
The filled circle marks the 1.3 mm continuum source L1641-N MM1
\citep{Stanke07}.
}
\label{fig11}
\end{figure}

Mapping observations in the \water\ line were carried out
on 2012 January 11 with a grid spacing of 16$''$.
The best-fit position of the source (KLC 4)
is $\sim$6$''$ west of L1641N MM1 (Figure \ref{fig11}).
We presume that KLC 4 is associated with MM1/3.
The \water\ maser reported by \cite{Xiang95} is displaced
by $\sim$40$''$ to the east with respect to KLC 4.
Considering their position uncertainty of $\lesssim$ 22$''$,
the displacement is significant.
It is not clear whether the maser of \cite{Xiang95} is excited by MM1/3
or by a different YSO in the L1641N cluster.

The \methanol\ 44 GHz line shows a double-peak profile
while the 95 GHz line shows a single peak (Figure \ref{fig10}). 
It is not clear whether the 44 GHz spectrum represents
two velocity components of maser emission
or a single component of thermal emission with a self-absorption feature.
We favor, however, the interpretation that the 44 GHz line consists of two narrow features.
The 95 GHz line shows a redshifted line wing.

The \methanol\ source position was determined by mapping the region
in the 95 GHz line with a grid spacing of 11$''$.
The best-fit source position of KLC 5
is ($-$16$''$, $-$10$''$) with respect to MM1 (Figure \ref{fig11}).
KLC 5 coincides within $\sim$3$''$
with the compact emission feature ``southwest shock''
traced by the \methanol\ $8 \rightarrow 7$ $E$ lines,
which is at the tip of the redshifted jet probably driven by MM3
\citep{Stanke07}.

\subsection{KLC 6/7 in the HOPS 361 Field (NGC 2071)}

Previous observations of the NGC 2071 region
showed many YSOs and various star formation activities
such as OH and \water\ masers \citep{Pankonin77,Genzel79,Tofani95},
infrared sources \citep{Persson81,Walther93},
compact radio sources \citep{Snell86,Torrelles98},
and molecular outflows \citep{Snell84,Scoville86,Choi93,Stojimirovi08}.
IRS 1 is the most luminous object in this region
and is an intermediate-mass class I protostar \citep{Skinner09}.
The distributions of radio continuum and \water\ maser emission suggest
that IRS 1 drives a jet in the east-west direction
\citep{Tofani95,Torrelles98,Trinidad09,Carrasco12}.
IRS 3 is a deeply embedded protostar,
but its nature is less certain \citep{Skinner09}.
IRS 3 drives a large-scale outflow/jet in the northeast-southwest direction,
and its \water\ masers seem to trace a rotating disk
\citep{Torrelles98,Eisloffel00,Trinidad09}.

\begin{figure}[!b]
\plotone{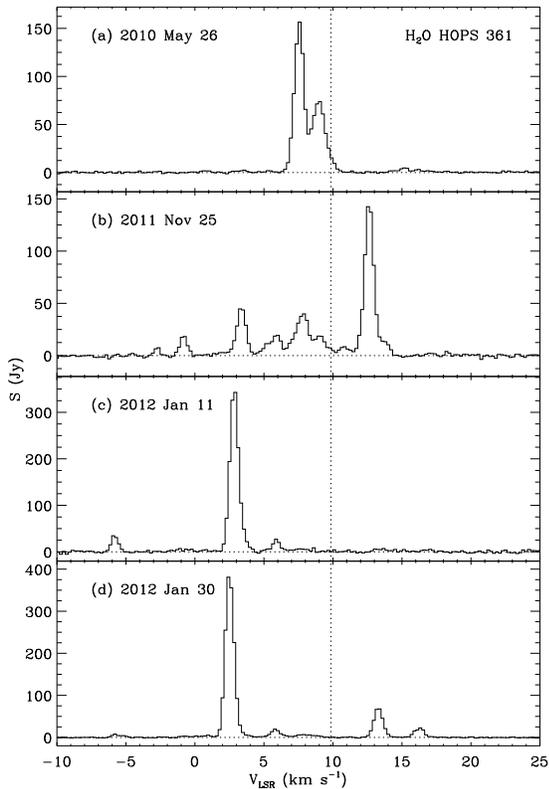}
\caption{
Spectra of the \water\ maser line toward HOPS 361 (KLC 6).
The vertical dotted line indicates
the systemic velocity of the ambient dense gas, $v_{\rm LSR}$ = 9.9 \kms\
\citep{Lada91}.
}
\label{fig12}
\end{figure}

\begin{figure}[!t]
\plotone{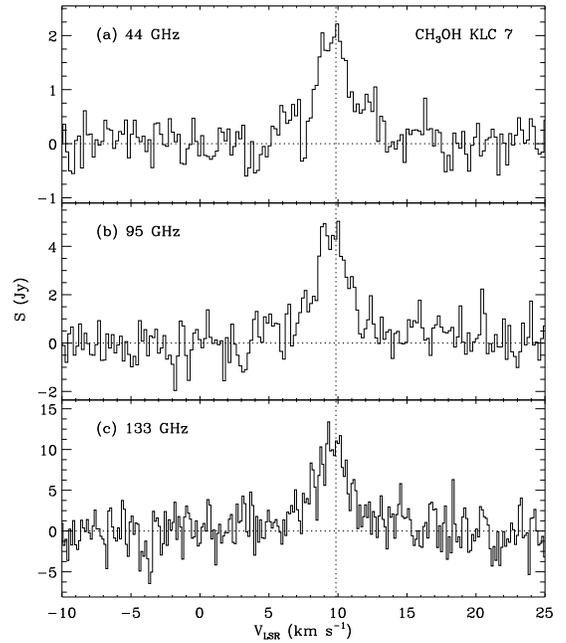}
\caption{
Spectra of the \methanol\ 44, 95, and 133 GHz lines
toward KLC 7 (NGC 2071 IRS 1).
The peak intensities of these lines in the main-beam temperature scale
are $T_{\rm mb}$ = 0.30, 0.72, and 1.4 K, respectively.
The vertical dotted line indicates the systemic velocity.
}
\label{fig13}
\end{figure}

The \water\ line and the three \methanol\ lines
were detected toward the NGC 2071 region (Figures \ref{fig12}--\ref{fig13}).
The \water\ maser spectrum displayed a strong variability
in flux and velocity over the four observing runs.
The velocity components were distributed
in a relatively wide range, from $-$6 to 16 \kms.

\begin{figure}[!t]
\plotone{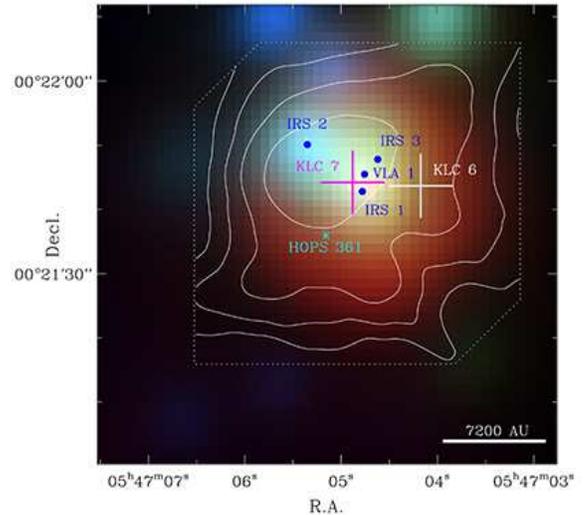}
\caption{
Positions of KLC 6/7 (white/magenta plus signs)
in the HOPS 361 (asterisk) field.
The filled circles mark the radio continuum sources in the NGC 2071 cluster
\citep{Snell86,Trinidad09}. The contours show the \methanol\ 95 GHz line map.
The line was integrated over the velocity interval of (6.0, 14.0) \kms.
The contour levels are 20\%, 40\%, 60\%, and 80\%\
of the maximum value (2.9 K \kms). The dotted polygon encloses the area covered.
}
\label{fig14}
\end{figure}

Mapping observations in the \water\ line were carried out
on 2011 November 25 with a grid spacing of 65$''$.
The source positions of all the velocity components detected on this day
agree to each other within the uncertainty.
The best-fit position of the source (KLC 6)
is $\sim$9$''$ west of IRS 1 (Figure \ref{fig14}). 
All the four radio sources in the NGC 2071 region (IRS 1--3 and VLA 1)
display \water\ maser activities \citep{Genzel79,Torrelles98,Trinidad09},  
but there is no previous report of maser at the position of KLC 6.
Interestingly, KLC 6 is at the intersection
of the IRS 1 western outflow and the IRS 3 southwestern outflow,
and it is difficult to point out the object
responsible for the excitation of the KLC 6 maser.
As the position difference between KLC 6 and the previously known
masers is less than 2 rms of pointing uncertainty, it is possible
that KLC 6 may not be a new maser.

The \methanol\ lines of KLC 7 (Figure \ref{fig13})
have a peak velocity close to the systemic velocity of the cloud
and show a width of $\sim$3.2 \kms,
larger than the line widths of KLC 1/5 by a factor of $\sim$4.
There is a possibility that the \methanol\
spectra of KLC 7 may have double peaks. In this case, the line may 
consist of a weak maser overlaid upon thermal emission.
The NGC 2071 region was mapped in the 95 GHz line
with a grid spacing of 10$''$.
The 95 GHz map shows that the source is extended
over a region of $\sim$50$''$ (Figure \ref{fig14}).
By contrast, \cite{Haschick89} found
that the \methanol\ 36 GHz maser line
has velocities redshifted by $\sim$5.5 \kms,
and \cite{Liechti96} reported
that the 36 GHz maser source is located $\sim$20$''$ south of IRS 1--3.
The large line width, extended source size,
and dissimilarity to the 36 GHz maser suggest
that the \methanol\ 44/95/133 GHz line emission of KLC 7 may be thermal.
The \methanol\ lines probably trace the dense molecular gas and outflows
in and around the NGC 2071 cluster \citep{Garay00}.

\subsection{KLC 8 in the HOPS 362 Field (V380 Ori NE)}

V380 Ori NE (HOPS 362) is a deeply embedded protostar \citep{Davis00}.
\cite{Zavagno97} classified it as a class I YSO,
but \cite{Stanke03} considered it class 0.
\cite{Stanke03} suggested that V380 Ori NE drives a molecular jet.
The bipolar outflow of V380 Ori NE
shows an interesting point-symmetric morphology in that 
the flow direction changes by 20\arcdeg\ when imaged in a larger scale
\citep{Davis00,Stanke03}.
\cite{Davis00} suggested that the jet is
either deflected by the ambient cloud
or driven by a precessing object.

\begin{figure}[!t]
\plotone{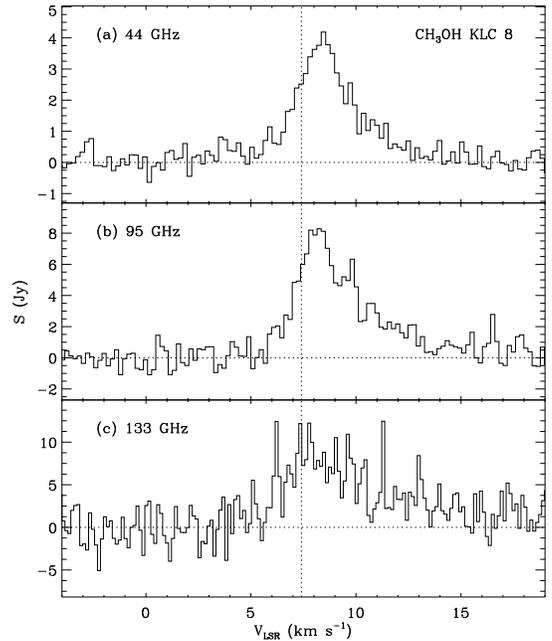}
\caption{
Spectra of the \methanol\ 44, 95, and 133 GHz lines
toward KLC 8 (V380 Ori NE S2).
The peak intensities in $T_{\rm mb}$ are 0.58, 1.1, and 1.1 K, respectively.
The vertical dotted line indicates
the systemic velocity of the ambient dense gas, $v_{\rm LSR}$ = 7.4 \kms\
\citep{Tatematsu98}.
}
\label{fig15}
\end{figure}

\begin{figure}[!t]
\plotone{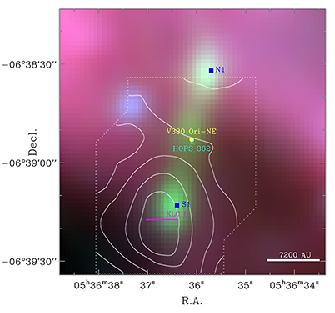}
\caption{
Position of KLC 8 (magenta plus sign) in the V380 Ori NE region.
The asterisk marks HOPS 362.
The filled circle marks the submillimeter continuum source,
and the filled squares mark the H$_2$ outflow knots \citep{Davis00}.
The contours show the \methanol\ 95 GHz line map. The line was integrated over the
velocity interval of (4.0, 12.0) \kms.
The contour levels are 20\%, 40\%, 60\%, and 80\%\
of the maximum value (5.3 K \kms). The dotted polygon encloses the area covered.}
\label{fig16}
\end{figure}

The three \methanol\ lines
were detected toward the V380 Ori NE region (Figure \ref{fig15}).
The \methanol\ lines of KLC 8
have a peak velocity redshifted by $\sim$1.2 \kms\
with respect to the systemic velocity of the cloud,
a line width of $\sim$4 \kms, and prominent redshifted line wings.
The V380 Ori NE region was mapped in the 95 GHz line
with a grid spacing of 10$''$.
The peak position is
near the H$_2$ line emission knots S$_1$--S$_3$ (Figure \ref{fig16}),
which is associated with the redshifted outflow lobe R$_1$
seen in the CO $J$ = 4 $\rightarrow$ 3 line \citep{Davis00}.
The 95 GHz map shows that the source is 
marginally resolved in the north-south direction
and unresolved in the east-west direction (Figure \ref{fig16}).
The map does not quite cover the northern (blueshifted) outflow area
but shows a hint of \methanol\ emission in that direction.
The large line width, extended source size, and the elongation suggest
that the \methanol\ line emission of KLC 8 may be thermal
and related to the southern outflow.

While the detection of \methanol\ emission
shows the presence of shocked gas near the bend of the outflow lobe,
it does not necessarily support or refute the deflection or precession
models \citep{Davis00}.
In the deflection scenario,
the \methanol\ emission may trace the dense gas deflecting the jet.
In the precession scenario,
it may trace the ambient cloud shocked freshly by the jet
carving a cavity in a new direction.

It is interesting to note
that the \methanol\ emission shows a good positional agreement
with the 4.6 \micron\ emission (Figure \ref{fig16}).
The \methanol\ emission in other regions (OMC 2, L1641N, and NGC 2071)
also shows a similar trend (Figures \ref{fig3}, \ref{fig11}, and \ref{fig14}).

\section{DISCUSSION}

Most of the stars in the Orion molecular cloud complex form in clusters.
This clustering should be taken into account
in the interpretation of the survey.
The targets were HOPS protostars showing CO line wings.
The CO observations, however, were made with a large beam,
and the detected CO line wings are not necessarily produced
by the outflows of the target protostars.
For example, the CO wings of HOPS 361
are most likely coming from the outflows driven by NGC 2071 IRS 1/3.
Therefore, the survey targets are not necessarily low-mass protostars
driving molecular outflows,
and they may rather be considered as typical star-forming regions
around low-mass protostars forming in clusters.

The clustering makes the interpretation of the single-dish survey complicated
because the main beam area can contain multiple protostars.
The consequence of clustering is obvious in the survey results.
Many of the detected masers are excited
not by the target HOPS protostars but by other YSOs in the survey fields.
For example, the exciting sources of KLC 2/3/6 are
HH 1--2 VLA 1, HH 1--2 VLA 3, and NGC 2071 IRS 1/3, respectively.
There are also some ambiguous cases.
Therefore, careful mapping is essential in single-dish surveys of masers.

While the effect of clustering in each region is difficult to predict,
we may assume that the degree of clustering
(i.e., the number of protostars in a main-beam area) in the survey regions
is more or less similar within an order of magnitude.
The reason is
that all the target sources in the survey are at similar distances
and that excessively complicated regions, such as the Orion KL region,
were excluded.

\subsection{\water\ Masers}

The nominal detection rate of the single-pointing \water\ maser survey
is 4\% (4/99).
The Orion KL masers hindered the detection of maser from nearby sources,
and the detection rate considering only the unaffected targets
is 5\% (3/60).
Since the mapping observations revealed
two maser sources in the HOPS 167 field,
the detection rate of the full survey is 7\% (4/60).
This value is somewhat smaller than the detection rate of 20--70\%  
from previous \water\ maser surveys of YSOs in various stages of evolution
\citep{Churchwell90,Furuya01,Sridharan02,Wilking94,Bae11}.
The relatively low rate from this survey is not surprising
because the targets are low-mass protostars.
Previous surveys are usually biased toward well-known (luminous) YSOs,
massive star-forming regions, or known maser sources.
For example, \cite{Furuya01} reported a detection rate of 
40\% for Class 0, 4\% for Class I, and 0\% for Class II low-mass YSOs. 
Their sample, however, consists of protostars 
known before the $Spitzer$ data became available. 
The results of this work is probably more representative
of typical low-mass star-forming regions.

In the regions described in Section 4,
a typical number of known protostars in a KVN main beam area is $\sim$4.
Then the detection rate,
defined as the number of detected masers per protostar in the covered region,
would be $\sim$1.7\%.
Another way to consider the clustering is to focus on the original target
sources only, not considering the off-center sources.
HOPS 182 (L1641N MM1/3) is the only detection among them,
and the detection rate would be 1.7\% (1/60).
Either way, the per-protostar detection rate is $\sim$1.7\%.
Therefore, an \water\ maser is a rare phenomenon in low-mass star formation.

Another factor complicating the interpretation is the time variability.
Some of the masers were detectable only at a certain epoch.
The maser in the HOPS 96 field was detected once in four observing runs.
KLC 2 was detected $\sim$4 times in five runs,
and KLC 3 was detected $\sim$2 times in five runs.
KLC 4/6 were detectable in all four runs.
Then the detection probability of these sources during a given run
is $\sim$70\%.
The detection rate (detections per protostar at a given epoch) of the survey
would be $\sim$1.2\%.

The \water\ maser is a good signpost of star formation activities,
especially shocked regions around the base of jets and on the disks.
Once detected, the \water\ maser is a useful tool
providing information on the mass accretion and ejection processes
on a small scale.
The rarity of \water\ masers, however, is one of the limitations of this tool
because $\sim$98\% of the protostars
would not exhibit detectable \water\ maser emission.
In other words, a non-detection of \water\ maser toward a known protostar
would not provide much constraints on the nature of the object.

One of the important properties of masers is
that the radiation can be anisotropic.
Since the \water\ masers in star-forming regions
are produced in structures that are far from spherical,
the detection of masers can depend
on the geometry of the emission/amplification region
\citep{Elitzur92}.
Therefore, the small detection rate above
means that \water\ masers are rarely detectable,
but does not necessarily imply that they are rarely occurring.

\subsection{\methanol\ Lines}

The survey resulted in the detection of four \methanol\ sources.
For a given source,
the three  \methanol\ lines used in the survey have similar spectral profiles,
which strongly suggests that they have the same origin.
Judging from several parameters measured
and from information in the literature,
KLC 1 in the OMC 2 region is clearly a maser source,
KLC 5 in L1641N is an ambiguous case,
and KLC 7/8 are thermal emission sources.
It is possible, however, that the
detected emission can be a mixture of maser and thermal components
to a certain degree \citep{Kalenskii10}.

The detection rate of \methanol\ class I masers is even smaller than
that of the \water\ masers. Since there are only one or two masers detected,
the nominal detection rate of the survey in the 44 GHz line is
1--2\%. The detection rates of \methanol\ masers toward high-mass
protostellar candidates range from 30\% to 50\%
\citep{Fontani10,Haschick90}.  It is clear that low-mass protostars
have a much smaller detection rate for \methanol\ class I masers.

Unlike \water\ masers, \methanol\ class I masers are often
detected far from protostars, which may be one of the reasons for
the low detection rate. Since the number of detections is so small,
it is difficult to make a further analysis. The effect of clustering
for the \methanol\ line is expected to be smaller than that for the
\water\ line because the beam size of the 44 GHz line is a factor of two 
smaller. KLC 1, however, is in a relatively crowded region (OMC 2
FIR 3/4) and probably not excited by the original target (HOPS 64).
Therefore, the per-protostar detection rate may be much smaller
than 1\%, and the \methanol\ class I maser is an even rarer phenomenon,
at least for low-mass protostars.

\section{SUMMARY}

Ninety-nine protostars in the Orion molecular cloud complex
were observed in the \water\ maser line at 22 GHz
and the \methanol\ class I maser lines at 44, 95, and 133 GHz
with the KVN antennas in its single-dish telescope mode.
The target sources are protostars
identified using the infrared observations
with the {\it Spitzer Space Telescope}.
The survey areas may be considered
as typical regions around low-mass protostars
at the same distance and in similar environments.
The main results are summarized as follows:

1. The \water\ maser line was detected toward four target sources
   (HOPS 96, 167, 182, and 361).
   The \water\ masers showed significant variability 
   in intensity and velocity on monthly timescales.

2. Regions around detected \water\ masers were mapped
   to identify the YSOs responsible for exciting the masers.
   KLC 2/3 in the HOPS 167 field may be excited
   by HH 1--2 VLA 3 and VLA 1, respectively.
   The VLA 1 \water\ maser is a new detection.
   KLC 4 in the HOPS 182 field may be excited by L1641N MM1/3.
   KLC 6 in the HOPS 361 field may be excited by NGC 2071 IRS 1/3.
   The maser in the HOPS 96 field may be excited by one of the YSOs
   in the OMC 3 SIMBA condensations,
   which is also a new detection.

3. The detection rate of \water\ masers,
   defined as number of detections per survey field,
   is 5--7\%.
   This value is lower than those of previous surveys
   probably because the targets in this survey are low-mass protostars.
   The detection rate, defined as detections per protostar,
   is $\sim$2\%.
   This small rate suggests
   that the \water\ maser of low-mass protostar
   is a rarely detectable phenomenon.

4. The \methanol\ 44, 95, and 133 GHz lines were detected
   toward four target sources (HOPS 64, 182, 361, and 362).
   The \methanol\ lines did not show significant variability
   and have peak velocities within $\sim$1 \kms\
   relative to the systemic velocities of the ambient dense clouds.
   The line width is a parameter
   useful for distinguishing maser and thermal emission,
   and the maser-thermal boundary is at $\sim$2 \kms.
   
5. Mapping observations in the 95 GHz line show
   that the detected \methanol\ sources are related to molecular outflows.
   KLC 1 in the HOPS 64 field is most likely a maser source.
   Its exciting source may be one of the protostars
   in the OMC 2 FIR 3/4 clusters.
   KLC 5 in the HOPS 182 field is probably a maser source,
   but interferometric observations are needed to verify its nature.
   It appears related to the jet driven by L1641N MM1/3.
   KLC 7/8 are probably thermal emission sources.
   KLC 7 in the HOPS 361 field is
   related with the dense cloud core and outflows
   in and around the NGC 2071 IRS 1--3 cluster.
   KLC 8 in the HOPS 362 field is related
   with the southern molecular outflow of V380 Ori NE.

6. The per-field detection rate of \methanol\ class I masers is 1--2\%.
   The per-protostar detection rate may be much smaller than 1\%.
   This small rate suggests
   that \methanol\ class I masers associated with low-mass protostar
   is an extremely rare phenomenon.

\acknowledgments 

We thank S. T. Megeath for a helpful discussion and  
the KVN staff for their support.
J.-E.L. was supported by the Basic Science Research Program
through the National Research Foundation of Korea (NRF)
funded by the Ministry of Education of the Korean government
(grant number NRF-2012R1A1A2044689) and the 2013 Sabbatical Leave
Program of Kyung Hee Unviersity (KHU-20131724).
M.C. was supported by the Core Research Program of NRF
funded by the Ministry of Science, ICT and Future Planning
of the Korean government (grant number NRF-2011-0015816).
This work was also supported
by the Korea Astronomy and Space Science Institute (KASI) grant
funded by the Korean government.


\end{document}